\documentclass[conference]{IEEEtran}
\IEEEoverridecommandlockouts
\usepackage{cite}
\usepackage{amsmath,amssymb,amsfonts}
\usepackage{algorithmic}
\usepackage{graphicx}
\usepackage{textcomp}
\usepackage{xcolor}
\usepackage{float}
\usepackage[hidelinks]{hyperref}
\usepackage{booktabs} % For formal tables
\usepackage{siunitx} % For aligning numbers by decimal point in tables
\usepackage{caption} % For captions in tables
\usepackage{adjustbox} % For adjusting box sizes

\def\BibTeX{{\rm B\kern-.05em{\sc i\kern-.025em b}\kern-.08em
    T\kern-.1667em\lower.7ex\hbox{E}\kern-.125emX}}

\begin{document}

\title{Evaluating BM3D and NBNet: A Comprehensive Study of Image Denoising Across Multiple Datasets}

\author{
\IEEEauthorblockN{Ghazal Kaviani, Reza Marzban, Ghassan AlRegib}
\IEEEauthorblockA{School of Electrical and Computer Engineering \\
Georgia Institute of Technology\\
Emails: gkaviani3@gatech.edu, mmarzban3@gatech.edu, alregib@gatech.edu}
}

\maketitle

\begin{abstract}
This paper investigates image denoising, comparing traditional non-learning-based techniques, represented by Block-Matching 3D (BM3D), with modern learning-based methods, exemplified by NBNet. We assess these approaches across diverse datasets, including CURE-OR, CURE-TSR, SSID+, Set-12, and Chest-Xray, each presenting unique noise challenges. Our analysis employs seven Image Quality Assessment (IQA) metrics and examines the impact on object detection performance. We find that while BM3D excels in scenarios like blur challenges, NBNet is more effective in complex noise environments such as under-exposure and over-exposure. The study reveals the strengths and limitations of each method, providing insights into the effectiveness of different denoising strategies in varied real-world applications.
\end{abstract}
\begin{IEEEkeywords}
Image denoising, non-local filtering, SHAP analysis, image quality assessment
\end{IEEEkeywords}

\section{Introduction}
% ... Your introduction here ...% ... Your related work here ...
In the field of image denoising, a multitude of techniques have been developed and refined over the years \cite{fan2019brief,goyal2020image}. In this study, we concentrate on two well-established categories of denoising methods: traditional non learning-based techniques, and more recent learning-based approaches. Traditional methods, such as Non-Local Means (NLM) \cite{liu2008robust} and Block-Matching 3D (BM3D) \cite{dabov2007image}, leverage local similarities within an image and the independent nature of noise. BM3D, one of the methods that we used here, works by searching for similar 2D image blocks within an image and stacking them in 3D arrays. This process takes advantage of the high level of similarity and redundancy in multidimensional data to effectively denoise images. However, this method is very resource consuming\cite{li2017improved} and the run-time exponentially grows with the input image size and quality. 
Another problem with traditional methods such as BM3D is they are not generalizable and for each noisy image or a set of similar noisy images, we need to tune model parameters to achieve the best results. More recent learning-based methods such as deep neural network (DNN)-based denoising strategies, utilize the implicit knowledge of image priors and noise characteristics, acquired through training on paired datasets\cite{zhang2017beyond}. While these convolutional neural network (CNN) approaches have seen substantial breakthroughs, they often struggle in complex scenarios, such as in areas with subtle textures or intricate high-frequency details. CNNs typically rely on local filters to differentiate between noise and actual signal. However, with a low signal-to-noise ratio (SNR), these local responses can become unreliable. In our study, we specifically investigate these effects by using the CURE-OR and CURE-TSR datasets which provide multiple levels of the same noise on images. Levels 3 -5 of each challenge in these datasets present a very low SNR. NBNEt\cite{cheng2021nbnet}, the other method that we used in our study, addresses this challenge by incorporating non-local image data through a technique called image projection. The network generates a set of basis vectors from the input image and reconstructs the image within a subspace defined by the span of these vectors. Given that natural images often exist within a low-rank signal subspace, reconstructing the image from this subspace should effectively suppress the noise. NBNet currently holds a state-of-the-art on some image-denoising benchmark. However, the power of both methods remains to be evaluated in this study.
\section{Datasets}
In this study, we systematically assess the performance of the denoising methods by applying them to a diverse range of datasets. Each dataset offers unique characteristics and challenges, ranging from varying noise levels to distinct image features and domains.
\subsubsection*{CURE-OR Dataset\cite{temel2018cure}} This dataset includes following noise challenges:
Underexposure, Overexposure, Blur, Contrast, Dirty lens, Image noise, Resizing, Decolorization 
\subsubsection*{CURE-TSR Dataset:\cite{Temel2017_NIPSW}}
This dataset includes the following noise challenges:
Snow, Haze, Rain, Shadow, Gaussian Blur, Exposure, Dirty lens, Darkening, Codec Error, Lens Blur, Decolorization 
\subsubsection*{SSID+ :\cite{abdelhamed2018high}}
SSID+ is a new version of SSID published as a challenge at the New Trends in Image Restoration and Enhancement (NTIRE 2020) workshop in conjunction with CVPR 2020.The dataset captures noisy images from 10 scenes under different lighting conditions using five representative smartphone cameras.
\subsubsection*{Set-12 :\cite{zhang2017beyond}}
The primary challenge associated with the Set-12 dataset is additive white Gaussian noise (AWGN), which is a standard model for noise in images.
\subsubsection*{Chest-Xray: \cite{kermany2018identifying}}
This dataset comprises X-ray images of patients categorized into two classes: normal and pneumonia-affected. A significant challenge presented by this dataset is the differential impact of Additive White Gaussian Noise (AWGN) on images from each class.

\section{Methodology}
This section focuses on explaining the two main denoising methods we used in our study: BM3D and NBNet. These methods were chosen because they represent two different types of denoising techniques. BM3D is known for its traditional approach, while NBNet uses a representation learning method. We will go into detail about how each of these methods works on selected datasets and the specific ways we set them up for our experiments.
\subsubsection*{BM3D} This method operates in two identical steps: initially generating a basic estimate of the noisy image using hard thresholding, followed by the actual denoising using the Wiener filter in the second step. For this purpose, BM3D uses the basic estimate from the first step as a reference or a pilot signal in the Wiener filter \cite{hasan2018improved}. BM3D struggles to generalize across different noise types, and its settings aren't automatically tailored to the input image's noise level. To optimize its performance, we used grid-search parameter tuning, where we tested various parameters on a subset of each dataset, challenges, and noise levels and selected the best ones based on the PSNR values of the denoised images. We then applied these optimal parameters to denoise the full set of images.\\
We also used MATLAB's parfor \cite{MATLAB} function to denoise images in parallel, randomly picking a certain number of images from each set. This was especially important for processing resource-intensive datasets like CURE-OR and CURE-TSR, which have various noise levels, and for large images like those in the SIDD dataset, since BM3D's computational load depends on the input image size.

\begin{figure}[h]
  \centering
  \begin{minipage}{0.4\textwidth}
    \centering
    \includegraphics[width=\textwidth]{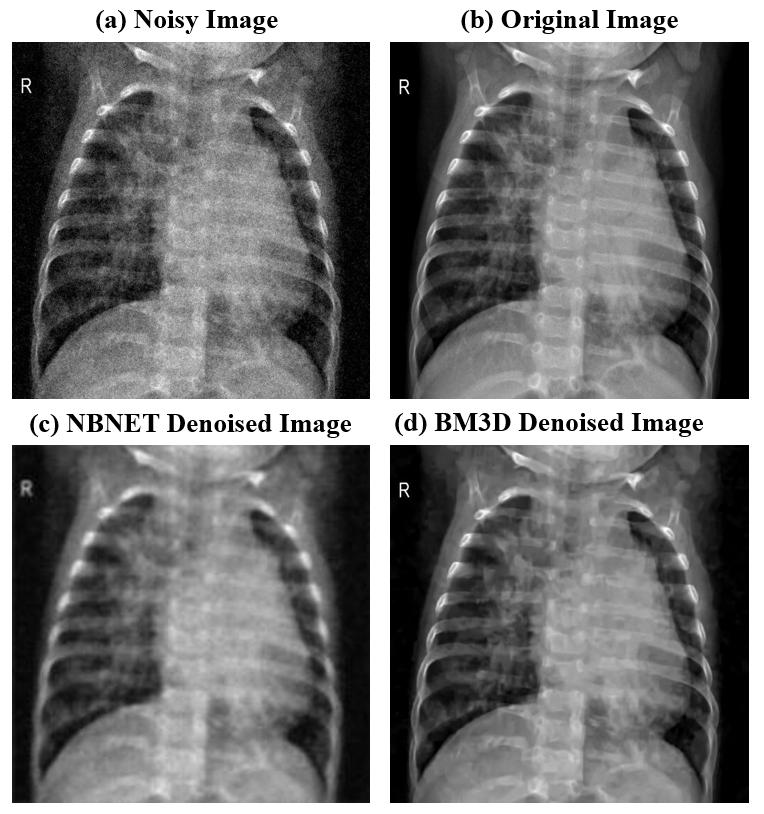}
    \captionsetup{font=scriptsize} % Set the font size of the caption
    \caption{Sample images from the Chest X-ray dataset depicting a case of pneumonia. The four panels include (a) the noisy input image, (b) the original image, (c) the image denoised using NBNet, and (d) the image denoised using BM3D.}
  \end{minipage}
\end{figure}

\subsubsection*{NBNet} is an adaptation of the U-Net structure designed for end-to-end image denoising via adaptive projection. It accomplishes denoising by selecting the appropriate basis of the signal subspace and projecting the input into this space through a non-local subspace attention module. This method is more adaptable to different noise types compared to BM3D. However, its basis vector and the subspace representation of the images are not universally applicable across various datasets. To demonstrate this, we used a pre-trained NBNet model on one dataset and then fine-tuned this model for another dataset. We observed that the denoising quality varied significantly across datasets with different noise challenges and levels. These differences are discussed in more detail in the Discussion section of the paper. 
\begin{figure}[h]
  \centering
  \begin{minipage}{0.4\textwidth}
    \centering
    \includegraphics[width=\textwidth]{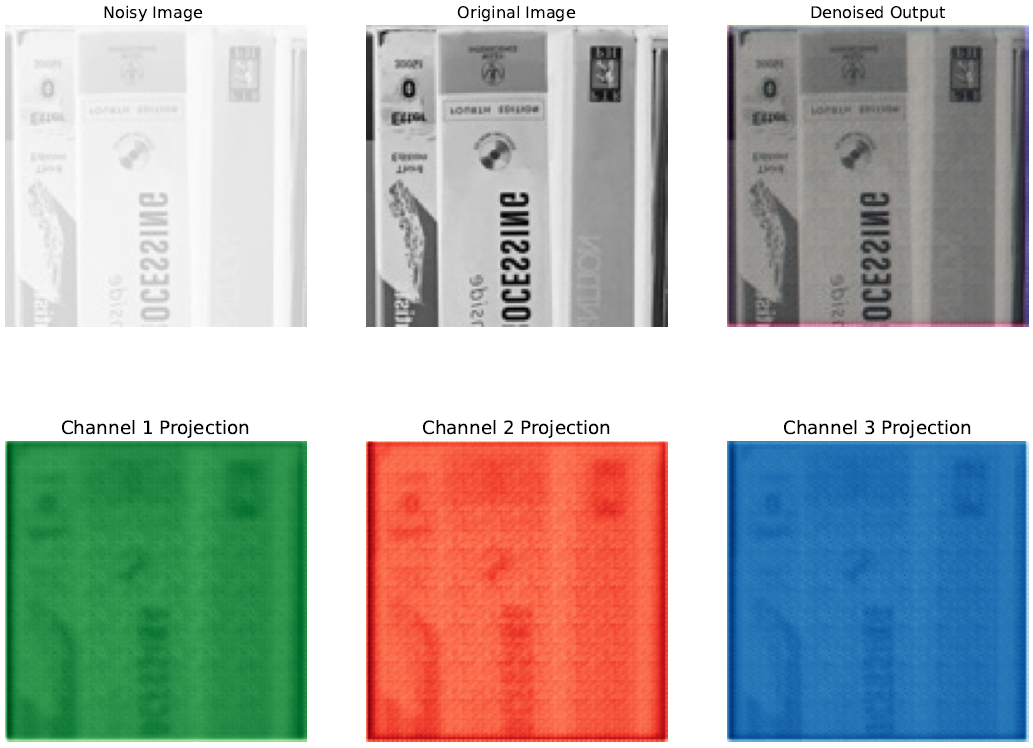}
        \captionsetup{font=scriptsize}
        \caption{NBNet employs subspace projection for denoising, generating basis for the signal subspace. This figure illustrates its effectiveness on level 3 noise from the CURE-OR dataset.}

  \end{minipage}
\end{figure}

\subsection*{IQA analysis} In this study, we assessed the performance of the denoising methods across all datasets using seven Image Quality Assessment (IQA) metrics: PSNR, SSIM, CW\_SSIM, UNIQUE\cite{temel2016unique}, MS\_UNIQUE\cite{prabhushankar2018ms}, CSV\cite{temel2016csv}, and SUMMER\cite{temel2019perceptual}. Additionally, for datasets with applications in detection or recognition, we incorporated a machine learning-based metric. This comprehensive approach allowed us to evaluate the denoising effectiveness not only through traditional IQA methods but also through a  machine-learning perspective on image quality and see how image quality impacts machine learning performance.
\section{Discussion}
In this section we discuss the outcomes of the different image denoising methods applied to the datasets. Highlight the performance of NBNET and BM3D methods in terms of the IQA metrics listed.
% \begin{figure}[h]
%   \centering
%   \begin{minipage}{0.45\textwidth}
%     \centering
%     \includegraphics[width=\textwidth]{Figs/learning_curve.PNG}
%         \caption{Training and validation loss curves for NBNet, illustrating the model's learning dynamics over epochs.}

%   \end{minipage}
% \end{figure}

% \begin{figure}[h]
%     \centering
%     \includegraphics[width=0.5\textwidth]{Figs/Linear Projection.png}
%     \caption{Transformation of seven distinct IQA metrics into a simplified representation through a linear projection.}
%     \label{fig:iqa-correlation-matrix}
% \end{figure}
% \begin{figure}[h]
%     \centering
%     \includegraphics[width=0.45\textwidth]{Figs/tsne.PNG}
%     \caption{2 dimensional projection of 7 IQA values }
%     \label{fig:iqa-correlation-matrix}
% \end{figure}

% \begin{figure}[htb]
%     \centering
%     \includegraphics[width=1\linewidth]{Figs/MS_UNIQUE.PNG}
%     \caption{SHAP Dependence Plot for the Feature \texttt{MS\_UNIQUE}}
%     \label{fig:shap_dependence_MS_UNIQUE}
% \end{figure}

Upon examining the IQA metrics for the SIDD dataset, it becomes clear that NBNet surpasses BM3D in performance. Metrics such as PSNR and SSIM, among others, show a twofold improvement in image quality, with the exception of Summer and CSV which are built for analyzing colored images. The NBNet model was thoroughly trained and validated on the extensive SIDD dataset. This trained model was subsequently used as a foundation for preprocessing other datasets.
% \begin{table*}[htbp]
\begin{table}[H]
% \captionsetup{justification=centering, singlelinecheck=true} % Uncomment this line if needed
\captionsetup{font=small}
\caption{\textbf{Summary of IQA Metrics for NBNET and BM3D Methods on the SIDD dataset}}
\label{tab:IQAMetrics}
% \centering
\large
\begin{adjustbox}{width=0.48\textwidth}
\begin{tabular}{@{}lS[table-format=2.2]S[table-format=1.3]S[table-format=1.3]S[table-format=1.3]S[table-format=1.3]S[table-format=1.3]S[table-format=1.3]@{}}
\toprule
\textbf{Method} & {\textbf{PSNR}} & {\textbf{SSIM}} & {\textbf{CW\_SSIM}} & {\textbf{UNIQUE}} & {\textbf{MS\_UNIQUE}} & {\textbf{CSV}} & {\textbf{SUMMER}} \\
\midrule
NBNET & 27.62 & 0.735 & 0.173 & 0.273 & 0.529 & 0.999 & 4.997 \\
\midrule
BM3D & 14.93 & 0.408 & 0.438 & 0.101 & 0.171 & 0.998 & 4.488 \\
\bottomrule
\end{tabular}
\end{adjustbox}
\end{table}

In the chest X-ray dataset, we encounter two distinct image classes, each impacted differently by additive noise. The pneumonia class, resembling AWGN (Additive White Gaussian Noise) in structure, presents a heightened challenge for denoising models. Contrasting with previous dataset observations, BM3D shows notably better performance in this context. This enhancement is primarily due to the noise's independence from the image structures. Although visual differences in denoised images by NBNet and BM3D are subtle, the PSNR and SSIM metrics indicate a substantial disparity in image quality. We hypothesize that metrics like CW\_SSIM, UNIQUE, and MS\_UNIQUE offer a more accurate reflection of image quality, particularly in structured images, as opposed to natural ones.

\begin{table}[H]
\scriptsize % Smaller font size
\small % or \footnotesize
\captionsetup{font=small}
\caption{\textbf{Summary of IQA Metrics Across Datasets and Methods Chest-Xray}}
\label{tab:IQAMetrics}
\begin{adjustbox}{width=0.48\textwidth}
\begin{tabular}{@{}lS[table-format=2.2]S[table-format=1.3]S[table-format=1.3]S[table-format=1.3]S[table-format=1.3]S[table-format=1.3]S[table-format=1.3]@{}}
\toprule
\textbf{Classes} & {\textbf{PSNR}} & {\textbf{SSIM}} & {\textbf{CW\_SSIM}} & {\textbf{UNIQUE}} & {\textbf{MS\_UNIQUE}} & {\textbf{CSV}} & {\textbf{SUMMER}} \\
\midrule
\multicolumn{8}{c}{\textbf{NBNET}} \\ % Centered bold label
\midrule
Normal & 23.41 & 0.755 & 0.462 & 0.357 & 0.614 & 0.999 & 4.965 \\
Pneumonia & 21.59 & 0.794 & 0.501 & 0.362 & 0.629 & 0.999 & 4.951 \\
\midrule
\multicolumn{8}{c}{\textbf{BM3D}} \\ % Centered bold label
\midrule
Normal & 33.04 & 0.874 & 0.528 & 0.567 & 0.762 & 1.000 & 4.984 \\
Pneumonia& 33.60 & 0.894 & 0.514 & 0.571 & 0.775 & 1.000 & 4.983 \\
\bottomrule
\end{tabular}
\end{adjustbox}
\end{table}

In assessing the IQA metrics for the Set-12 dataset, it's clear that the BM3D method outperforms the pre-trained NBNet model in denoising, particularly by looking at the UNIQUE and MS\_UNIQUE metrics.The CSV and SUMMER scores are identical for both methods, likely due to the use of grayscale images. Set-12 only contains 12 images which is far less than sufficient for fine-tuning a model. 

\begin{table}[H]
\scriptsize % Smaller font size
\small % or \footnotesize
\captionsetup{font=small}
\caption{\textbf{Summary of IQA Metrics for NBNET and BM3D Methods set 12}}
\label{tab:IQAMetrics}
\begin{adjustbox}{width=0.48\textwidth}
\begin{tabular}{@{}lS[table-format=2.2]S[table-format=1.3]S[table-format=1.3]S[table-format=1.3]S[table-format=1.3]S[table-format=1.3]S[table-format=1.3]@{}}
\toprule
\textbf{Method} & {\textbf{PSNR}} & {\textbf{SSIM}} & {\textbf{CW\_SSIM}} & {\textbf{UNIQUE}} & {\textbf{MS\_UNIQUE}} & {\textbf{CSV}} & {\textbf{SUMMER}} \\
\midrule
NBNET &  21.901 & 0.633 & 0.393 & 0.175 & 0.398 & 0.998 & 4.974 \\
\midrule
BM3D &  26.864 & 0.764 & 0.412 & 0.493 & 0.639 & 0.999 & 4.990\\
\bottomrule
\end{tabular}
\end{adjustbox}
\end{table}

For the CURE-OR dataset, we've added an extra evaluation from a machine learning standpoint. We utilized a Faster RCNN architecture, as outlined in Ren et al. (2015)\cite{ren2015faster}, with weights pre-trained on the COCO object dataset \cite{lin2014microsoft}. We used the original images as a baseline for reference. Then, we tested the model on images denoised by both methods, measuring performance using an F-1 score derived from precision and recall. This score compares the original and denoised images object detection. Due to the range of noise challenges and the very low SNR in levels 3 to 5 of the CURE-OR dataset, detection performance is notably low—almost impossible in some cases. Training an object detection model from scratch on these noisy images for a sufficient number of epochs could yield much better results by learning the relevant noise distributions. However, given the scope and time constraints of this project, we haven't been able to train the model on all subsets of this dataset. The methods show varying performances across the different challenges in this dataset which we discussed one by one in the following.

BM3D outperformed NBNet in denoising images in the Blur challenges, as indicated by all IQA metrics except PSNR. This assessment is further supported by the object detection model's ability to identify objects in images denoised by BM3D at level 1 of the challenge, suggesting that metrics other than PSNR provide a more accurate evaluation in this context.

\begin{table}[H]
\scriptsize % Smaller font size
\small % or \footnotesize
\captionsetup{font=small}
\caption{\textbf{Summary of IQA Metrics for CURE-OR Dataset Blur Challenge}}
\begin{adjustbox}{width=0.48\textwidth}
\begin{tabular}{@{}lS[table-format=2.6]S[table-format=1.3]S[table-format=1.3]S[table-format=1.3]S[table-format=1.3]S[table-format=1.3]S[table-format=1.3]S[table-format=1.0]@{}}
\toprule
\textbf{Noise level} & {\textbf{PSNR}} & {\textbf{SSIM}} & {\textbf{CW\_SSIM}} & {\textbf{UNIQUE}} & {\textbf{MS\_UNIQUE}} & {\textbf{CSV}} & {\textbf{SUMMER}} & {\textbf{F1 Score}} \\

\midrule
\multicolumn{9}{c}{\textbf{NBNET}} \\ % Centered bold label
\midrule

\textbf{Level 1 }& 18.347 & 0.677 & 0.359 & 0.293 & 0.488& 0.998 & 4.983 & 0  \\
\textbf{Level 2 }& 16.255 & 0.642 & 0.330 & 0.161 & 0.385 & 0.997 & 4.994 & 0 \\
\textbf{Level 3 }& 15.026& 0.626 & 0.317& 0.112 & 0.342 & 0.997 & 4.995 & 0 \\
\textbf{Level 4 }& 14.536 & 0.621& 0.309 & 0.088 & 0.314 & 0.997 & 4.997 & 0  \\
\textbf{Level 5 }& 14.208 & 0.616 & 0.299& 0.072 & 0.294 & 0.997 & 4.997 & 0  \\
\midrule
\multicolumn{9}{c}{\textbf{BM3D}} \\ % Centered bold label
\midrule

\textbf{Level 1}& 14.070 & 0.765 & 0.978& 0.474& 0.625& 0.998& 4.978 & 0.266 \\
\textbf{Level 2}& 13.739 & 0.740 & 0.526& 0.288& 0.494& 0.998& 4.993 & 0 \\
\textbf{Level 3}& 13.495 & 0.729 & 0.181& 0.220& 0.441& 0.998& 4.996& 0 \\
\textbf{Level 4}& 13.309 & 0.721 & 0.121& 0.180& 0.405& 0.997& 4.997& 0 \\
\textbf{Level 5}& 13.156 & 0.714 & 0.110& 0.153& 0.378& 0.997& 4.998& 0 \\
\bottomrule
\end{tabular}
\end{adjustbox}
\end{table}

In the resize challenge, the NBNet model shows a notable side effect when denoising images. This challenge doesn't introduce noise separate from the signal, but the pre-trained NBNet model reconstructs images with artifacts that significantly reduce their quality, as revealed by the IQA metrics. In contrast, BM3D manages to restore images with minimal artifacts. Moreover, the F-1 scores for images denoised by both methods indicate that the image quality is sufficient for machine-based object detection.

\begin{table}[H]
\scriptsize % Smaller font size
\captionsetup{font=small}
\caption{\textbf{Summary of IQA Metrics for CURE-OR Dataset Resize Challenge}}
\label{tab:CUREORGrayscaleResize}
\begin{adjustbox}{width=0.5\textwidth}
\begin{tabular}{@{}lS[table-format=2.6]S[table-format=1.3]S[table-format=1.3]S[table-format=1.3]S[table-format=1.3]S[table-format=1.3]S[table-format=1.3]S[table-format=1.0]@{}}
\toprule
\textbf{Noise level} & {\textbf{PSNR}} & {\textbf{SSIM}} & {\textbf{CW\_SSIM}} & {\textbf{UNIQUE}} & {\textbf{MS\_UNIQUE}} & {\textbf{CSV}} & {\textbf{SUMMER}} & {\textbf{F1 Score}} \\
\midrule
\multicolumn{8}{c}{\textbf{NBNET}} \\ % Centered bold label
\midrule
\textbf{Level 1}& 20.269 & 0.753& 0.543 & 0.445 & 0.577 & 0.998 & 4.977 & 0.659 \\
\textbf{Level 2}& 14.015 & 0.633& 0.449 & 0.359 & 0.514 & 0.998 & 4.998 & 0 \\
\textbf{Level 3}& 11.151 & 0.577& 0.395 & 0.254 & 0.444 & 0.997 & 4.999 & 0 \\
\textbf{Level 4}& 12.847 & 0.616& 0.452 & 0.359 & 0.508 & 0.998 & 4.999 & 0 \\
\midrule
\multicolumn{8}{c}{\textbf{BM3D}} \\ % Centered bold label
\midrule
\textbf{Level 1}& 69.233 & 0.999 & 0.976& 0.987 & 0.998 & 0.999 & 4.999 & 0.721 \\
\textbf{Level 2}& 68.697 & 0.999 & 0.970& 0.985 & 0.998 & 0.999 & 4.999 & 0 \\
\textbf{Level 3}& 68.130 & 0.999 & 0.963& 0.981 & 0.997 & 0.999 & 4.999 & 0 \\
\textbf{Level 4}& 67.388 & 0.998 & 0.951& 0.975 & 0.997 & 0.999 & 4.999 & 0 \\
\bottomrule
\end{tabular}
\end{adjustbox}
\end{table}

\begin{figure}[H]
    \centering
    \includegraphics[width=0.45\textwidth]{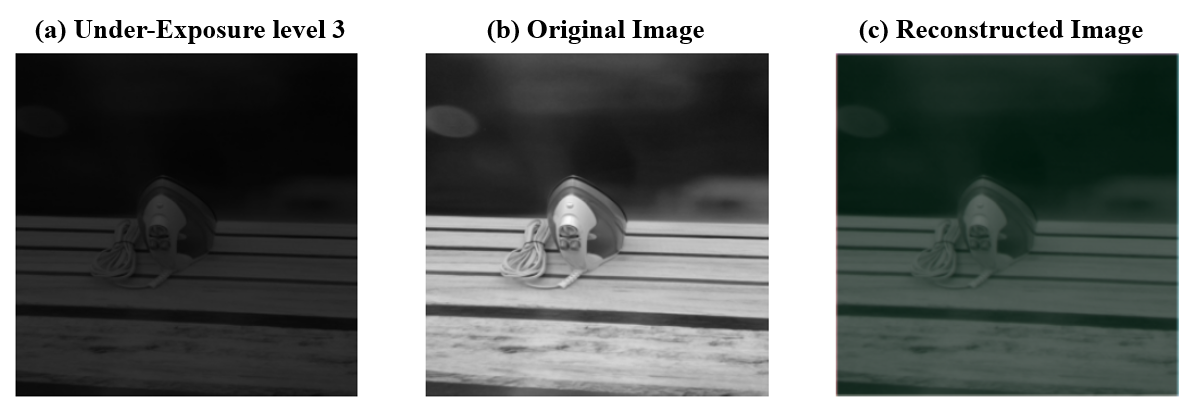}
     \captionsetup{font=scriptsize}
    \caption{Sample image denoised by NBNet model}
    \label{fig:Under-Expo level 3}
\end{figure}
In the under-exposure and over-exposure challenges, where the noise is not independent of the signal, BM3D struggles to effectively filter it out at any stage of denoising. Conversely, the NBNet model successfully reconstructs acceptable images from the subspace \ref{fig:Under-Expo level 3}, a performance supported by both the IQA metrics and F-1 scores.
\begin{table}[H]
\scriptsize % Smaller font size
\small % or \footnotesize
\captionsetup{font=small}
\caption{\textbf{Summary of IQA Metrics for CURE-OR Dataset Underexposure Challenge}}
\label{tab:CUREORGrayscaleUnderexposure}
\begin{adjustbox}{width=0.48\textwidth}
\begin{tabular}{@{}lS[table-format=2.6]S[table-format=1.3]S[table-format=1.3]S[table-format=1.3]S[table-format=1.3]S[table-format=1.3]S[table-format=1.3]S[table-format=1.0]@{}}
\toprule
\textbf{Noise level} & {\textbf{PSNR}} & {\textbf{SSIM}} & {\textbf{CW\_SSIM}} & {\textbf{UNIQUE}} & {\textbf{MS\_UNIQUE}} & {\textbf{CSV}} & {\textbf{SUMMER}} & {\textbf{F1 Score}} \\
\midrule
\multicolumn{8}{c}{\textbf{NBNET}} \\ % Centered bold label
\midrule
\textbf{Level 1}& 10.739 & 0.582 & 0.412 & 0.448 & 0.594 & 0.998 & 4.998 & 0.334 \\
\textbf{Level 2}& 9.598 & 0.496 & 0.366 & 0.365 & 0.534 & 0.998 & 4.999  & 0 \\
\textbf{Level 3}& 9.311 & 0.504 & 0.400 & 0.374 & 0.536 & 0.998 & 4.999  & 0 \\
\textbf{Level 4}& 11.174 & 0.641 & 0.428 & 0.402 & 0.563 & 0.998 & 4.999 & 0 \\
\textbf{Level 5}& 10.212 & 0.618 & 0.387 & 0.373 & 0.538 & 0.998 & 4.999 & 0 \\
\midrule
\multicolumn{8}{c}{\textbf{BM3D}} \\ % Centered bold label
\midrule
\textbf{Level 1 }& 4.928 & 0.009 & 0.019 & 0.017 & 0.154 & 0.998 & 5.000 & 0 \\
\textbf{Level 2 }& 4.915 & 0.008 & 0.016 & 0.015 & 0.149 & 0.998 & 5.000 & 0 \\
\textbf{Level 3 }& 4.895 & 0.005 & 0.007 & 0.002 & 0.104 & 0.998 & 5.000 & 0 \\
\textbf{Level 4 }& 4.883 & 0.004 & 0.000 & 0.000 & 0.084 & 0.998 & 5.000 & 0 \\
\textbf{Level 5 }& 4.887 & 0.004 & 0.000 & 0.000 & 0.084 & 0.998 & 5.000 & 0 \\
\bottomrule
\end{tabular}
\end{adjustbox}
\end{table}

In the DirtlyLens1, DirtyLens2, and Salt \& Pepper challenges of the CURE-OR dataset, which feature similarly structured noises, a consistent pattern emerges in the denoised images. While the PSNR, SSIM, and CW\_SSIM values are relatively close for both BM3D and NBNet methods, the UNIQUE and MS\_UNIQUE metrics reveal a better quality enhancement for BM3D. This gap is corroborated by the F-1 scores, which indicate that BM3D is more effective in denoising images for machine-based object detection.

\begin{table}[H]
\centering
\scriptsize % Smaller font size
\captionsetup{font=small}
\caption{\textbf{Summary of IQA Metrics for CURE-OR Dataset Dirtylens 1 Challenge  }}
\label{tab:YourDatasetIQA}
\begin{adjustbox}{width=0.48\textwidth}
\begin{tabular}{@{}lS[table-format=2.6]S[table-format=1.3]S[table-format=1.3]S[table-format=1.3]S[table-format=1.3]S[table-format=1.3]S[table-format=1.3]S[table-format=1.0]@{}}
\toprule
\textbf{Noise level} & {\textbf{PSNR}} & {\textbf{SSIM}} & {\textbf{CW\_SSIM}} & {\textbf{UNIQUE}} & {\textbf{MS\_UNIQUE}} & {\textbf{CSV}} & {\textbf{SUMMER}} & {\textbf{F1 Score}} \\
\midrule
\multicolumn{8}{c}{\textbf{NBNET}} \\ % Centered bold label
\midrule
\textbf{Level 1}& 21.043& 0.709 & 0.642 & 0.298 & 0.485& 0.998 & 4.934& 0 \\
\textbf{Level 2}& 18.892& 0.608 & 0.632 & 0.166 & 0.362& 0.998 & 4.850& 0 \\
\textbf{Level 3}& 15.360& 0.559 & 0.614 & 0.105 & 0.300& 0.998 & 4.797& 0 \\
\textbf{Level 4}& 14.607& 0.453 & 0.604 & 0.038 & 0.185& 0.997 & 4.499& 0 \\
\textbf{Level 5}& 12.554& 0.549 & 0.444 & 0.011 & 0.140& 0.997 & 4.886& 0 \\
\midrule
\multicolumn{8}{c}{\textbf{BM3D}} \\ % Centered bold label
\midrule
\textbf{Level 1}& 22.942 & 0.866 & 0.750 & 0.531& 0.714& 0.998 & 4.972 & 0.461 \\
\textbf{Level 2}& 19.434 & 0.814 & 0.644 & 0.317& 0.549& 0.998 & 4.900 & 0 \\
\textbf{Level 3}& 16.899 & 0.749 & 0.553 & 0.158& 0.384& 0.998 & 4.790 & 0 \\
\textbf{Level 4}& 14.912 & 0.680 & 0.470 & 0.061& 0.244& 0.997 & 4.637 & 0 \\
\textbf{Level 5}& 13.311 & 0.616 & 0.395 & 0.015& 0.139& 0.997 & 4.455 & 0 \\
\bottomrule
\end{tabular}
\end{adjustbox}
\end{table}

\begin{table}[H]
\centering
\scriptsize % Smaller font size
\captionsetup{font=small}
\caption{\textbf{Summary of IQA Metrics for CURE-OR Dataset Dirtylens 2 Challenge  }}

\label{tab:GrayscaleDirtyLens2}
\begin{adjustbox}{width=0.48\textwidth}
\begin{tabular}{@{}lS[table-format=2.6]S[table-format=1.3]S[table-format=1.3]S[table-format=1.3]S[table-format=1.3]S[table-format=1.3]S[table-format=1.3]S[table-format=1.0]@{}}
\toprule
\textbf{Noise level} & {\textbf{PSNR}} & {\textbf{SSIM}} & {\textbf{CW\_SSIM}} & {\textbf{UNIQUE}} & {\textbf{MS\_UNIQUE}} & {\textbf{CSV}} & {\textbf{SUMMER}} & {\textbf{F1 Score}} \\
\midrule
\multicolumn{8}{c}{\textbf{NBNET}} \\ % Centered bold label
\midrule
\textbf{Level 1}& 18.424 & 0.662 & 0.586 & 0.200& 0.378 & 0.998 & 4.731 & 0 \\
\textbf{Level 2}& 17.743 & 0.612 & 0.589 & 0.151& 0.324 & 0.998 & 4.675 & 0 \\
\textbf{Level 3}& 15.636 & 0.562 & 0.584 & 0.144& 0.315 & 0.998 & 4.612 & 0 \\
\textbf{Level 4}& 12.728 & 0.512 & 0.558 & 0.098& 0.264 & 0.997 & 4.704 & 0 \\
\textbf{Level 5}& 10.979 & 0.432 & 0.539 & 0.059& 0.213 & 0.997 & 4.624 & 0 \\
\midrule
\multicolumn{8}{c}{\textbf{BM3D}} \\ % Centered bold label
\midrule
\textbf{Level 1}& 20.139 & 0.852& 0.734 & 0.323 & 0.525 & 0.998 & 4.753& 0.377 \\
\textbf{Level 2}& 16.532 & 0.793& 0.647 & 0.213 & 0.418 & 0.998 & 4.619& 0 \\
\textbf{Level 3}& 13.970 & 0.736& 0.598 & 0.179 & 0.393 & 0.998 & 4.634& 0 \\
\textbf{Level 4}& 12.180 & 0.684& 0.545 & 0.121 & 0.326 & 0.997 & 4.636& 0 \\
\textbf{Level 5}& 10.010 & 0.573& 0.470 & 0.074 & 0.255 & 0.997 & 4.620& 0 \\
\bottomrule
\end{tabular}
\end{adjustbox}
\end{table}

\subsection*{IQA SHAP Analysis}
Evaluating seven IQA values across various image sets is a complex and time-consuming task, particularly when minor differences and inconsistencies among the metrics complicate reaching a clear conclusion. This complexity prompts an in-depth analysis of each metric's contribution to overall image quality assessment, which is crucial for assessing the effectiveness of different denoising methods on large datasets.
\begin{figure}[H]
    \centering
    \includegraphics[width=0.4\textwidth]{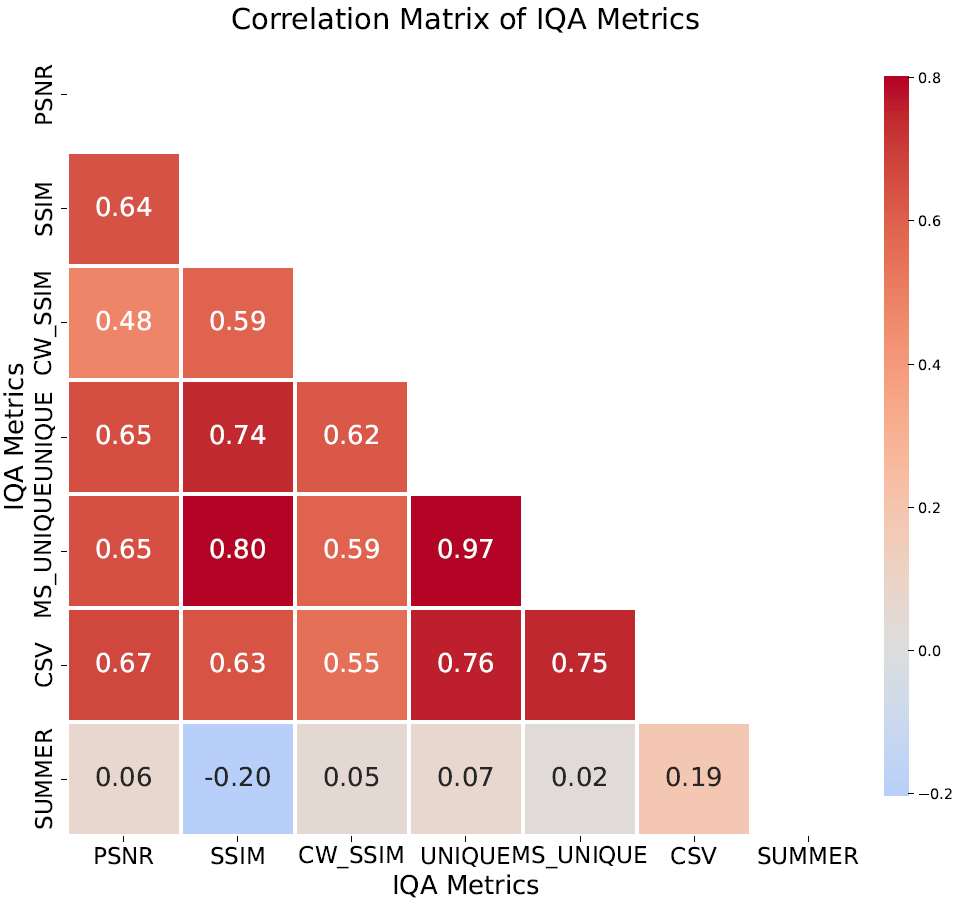}
     \captionsetup{font=scriptsize}
    \caption{Correlation Matrix of IQA Metrics. The heatmap illustrates the correlation between different Image Quality Assessment (IQA) metrics. Each cell displays the correlation coefficient, with warmer colors indicating stronger correlations. A larger font size is used for better readability, and the color bar numbers are also enlarged for clarity.}
    \label{fig:iqa-correlation-matrix}
\end{figure}
\subsubsection*{SHAP Analysis and Feature Importance}
Looking at figure \ref{fig:iqa-correlation-matrix}
we saw that there is a strong correlation between some of these values thus We employed SHAP analysis \cite{cohen2005feature} utilizing the DeepExplainer from the SHAP library to compute values reflecting each feature's contribution. Feature importance was gauged through mean absolute SHAP values across the test datasets.

\begin{figure}[H]
  \centering
  \begin{minipage}{0.48\textwidth}
    \centering
    \includegraphics[width=\textwidth]{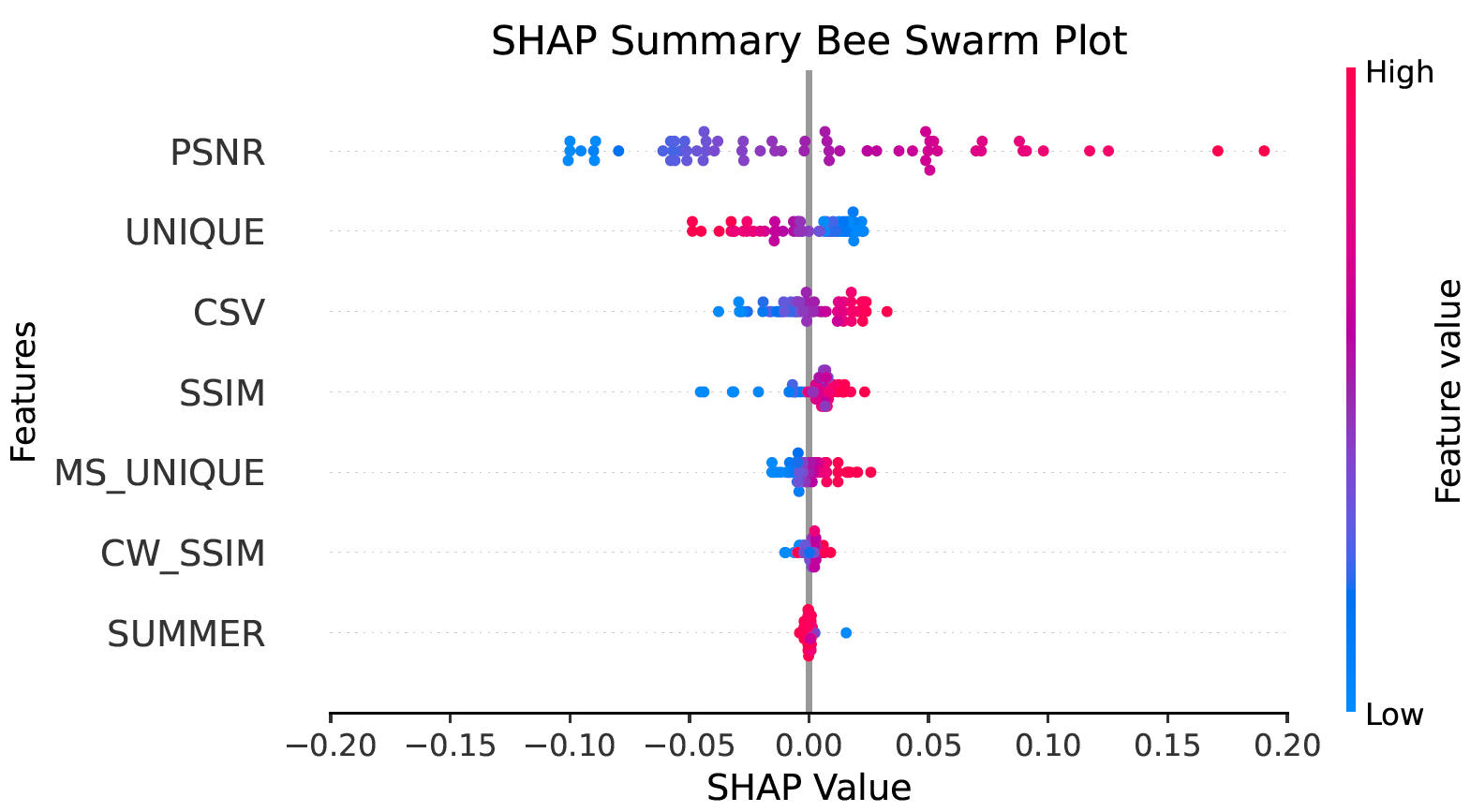}
    \captionsetup{font=scriptsize}
    \caption{SHAP Summary Bee Swarm and Bar Plot showing feature distribution and importance in the model.}
    \label{fig:shap-summary} % Adding the label here
  \end{minipage}
\end{figure}

\subsubsection*{Conclusion}

As illustrated in Figure \ref{fig:shap-summary}, the PSNR metric, along with UNIQUE and CSV, play a significant role in evaluating various denoising methods. It's important to note that these findings are specific to our test datasets. The relatively low score for the Summer metric is due to the lack of color-distorted samples in these datasets\\
\textbf{Rest of the analysis results for CURE-OR and CURE-TSR are in Appendix A after references.}
\clearpage % Add this line to start a new page before the bibliography
% \bibliographystyle{plain}
% \bibliography{references} % Use the name of your .bib file without the extension
\bibliographystyle{IEEEtran}
\bibliography{references} % references.bib is your bibliography file

@article{temel2016unique,
  title={UNIQUE: Unsupervised image quality estimation},
  author={Temel, Dogancan and Prabhushankar, Mohit and AlRegib, Ghassan},
  journal={IEEE signal processing letters},
  volume={23},
  number={10},
  pages={1414--1418},
  year={2016},
  publisher={IEEE}
}

@article{cohen2005feature,
  title={Feature selection based on the shapley value},
  author={Cohen, Shay and Ruppin, Eytan and Dror, Gideon},
  journal={other words},
  volume={1},
  pages={98Eqr},
  year={2005}
}

@article{prabhushankar2018ms,
  title={Ms-unique: Multi-model and sharpness-weighted unsupervised image quality estimation},
  author={Prabhushankar, Mohit and Temel, Dogancan and AlRegib, Ghassan},
  journal={arXiv preprint arXiv:1811.08947},
  year={2018}
}

@inproceedings{lin2014microsoft,
  title={Microsoft coco: Common objects in context},
  author={Lin, Tsung-Yi and Maire, Michael and Belongie, Serge and Hays, James and Perona, Pietro and Ramanan, Deva and Doll{\'a}r, Piotr and Zitnick, C Lawrence},
  booktitle={Computer Vision--ECCV 2014: 13th European Conference, Zurich, Switzerland, September 6-12, 2014, Proceedings, Part V 13},
  pages={740--755},
  year={2014},
  organization={Springer}
}

@article{ren2015faster,
  title={Faster r-cnn: Towards real-time object detection with region proposal networks},
  author={Ren, Shaoqing and He, Kaiming and Girshick, Ross and Sun, Jian},
  journal={Advances in neural information processing systems},
  volume={28},
  year={2015}
}

@article{temel2016csv,
  title={CSV: Image quality assessment based on color, structure, and visual system},
  author={Temel, Dogancan and AlRegib, Ghassan},
  journal={Signal Processing: Image Communication},
  volume={48},
  pages={92--103},
  year={2016},
  publisher={Elsevier}
}

@article{temel2019perceptual,
  title={Perceptual image quality assessment through spectral analysis of error representations},
  author={Temel, Dogancan and AlRegib, Ghassan},
  journal={Signal Processing: Image Communication},
  volume={70},
  pages={37--46},
  year={2019},
  publisher={Elsevier}
}

@article{goyal2020image,
  title={Image denoising review: From classical to state-of-the-art approaches},
  author={Goyal, Bhawna and Dogra, Ayush and Agrawal, Sunil and Sohi, Balwinder Singh and Sharma, Apoorav},
  journal={Information fusion},
  volume={55},
  pages={220--244},
  year={2020},
  publisher={Elsevier}
}

@article{li2017improved,
  title={Improved BM3D denoising method},
  author={Li, YingJiang and Zhang, Jiangwei and Wang, Maoning},
  journal={IET Image Processing},
  volume={11},
  number={12},
  pages={1197--1204},
  year={2017},
  publisher={Wiley Online Library}
}

@software{MATLAB,
year = {2022},
author = {The MathWorks Inc.},
title = {MATLAB version: 9.13.0 (R2022b)},
publisher = {The MathWorks Inc.},
address = {Natick, Massachusetts, United States},
url = {https://www.mathworks.com}
}

@article{zhang2017beyond,
  title={Beyond a gaussian denoiser: Residual learning of deep cnn for image denoising},
  author={Zhang, Kai and Zuo, Wangmeng and Chen, Yunjin and Meng, Deyu and Zhang, Lei},
  journal={IEEE transactions on image processing},
  volume={26},
  number={7},
  pages={3142--3155},
  year={2017},
  publisher={IEEE}
}

@article{fan2019brief,
  title={Brief review of image denoising techniques},
  author={Fan, Linwei and Zhang, Fan and Fan, Hui and Zhang, Caiming},
  journal={Visual Computing for Industry, Biomedicine, and Art},
  volume={2},
  pages={1--12},
  year={2019},
  publisher={Springer}
}

@article{liu2008robust,
  title={A robust and fast non-local means algorithm for image denoising},
  author={Liu, Yan-Li and Wang, Jin and Chen, Xi and Guo, Yan-Wen and Peng, Qun-Sheng},
  journal={Journal of computer science and technology},
  volume={23},
  number={2},
  pages={270--279},
  year={2008},
  publisher={Springer}
}

@inproceedings{cheng2021nbnet,
  title={Nbnet: Noise basis learning for image denoising with subspace projection},
  author={Cheng, Shen and Wang, Yuzhi and Huang, Haibin and Liu, Donghao and Fan, Haoqiang and Liu, Shuaicheng},
  booktitle={Proceedings of the IEEE/CVF conference on computer vision and pattern recognition},
  pages={4896--4906},
  year={2021}
}

@article{dabov2007image,
  title={Image denoising by sparse 3-D transform-domain collaborative filtering},
  author={Dabov, Kostadin and Foi, Alessandro and Katkovnik, Vladimir and Egiazarian, Karen},
  journal={IEEE Transactions on image processing},
  volume={16},
  number={8},
  pages={2080--2095},
  year={2007},
  publisher={IEEE}
}

@inproceedings{temel2018cure,
  title={Cure-or: Challenging unreal and real environments for object recognition},
  author={Temel, Dogancan and Lee, Jinsol and AlRegib, Ghassan},
  booktitle={2018 17th IEEE international conference on machine learning and applications (ICMLA)},
  pages={137--144},
  year={2018},
  organization={IEEE}
}

@INPROCEEDINGS{Temel2017_NIPSW,
Author = {D. Temel and G. Kwon and M. Prabhushankar and G. AlRegib},
Title = {{CURE-TSR: Challenging unreal and real environments for traffic sign recognition}},
Year = {2017},
booktitle = {Neural Information Processing Systems (NeurIPS) Workshop on Machine Learning for Intelligent Transportation Systems}}

@inproceedings{abdelhamed2018high,
  title={A high-quality denoising dataset for smartphone cameras},
  author={Abdelhamed, Abdelrahman and Lin, Stephen and Brown, Michael S},
  booktitle={Proceedings of the IEEE conference on computer vision and pattern recognition},
  pages={1692--1700},
  year={2018}
}

@article{hasan2018improved,
  title={Improved BM3D image denoising using SSIM-optimized Wiener filter},
  author={Hasan, Mahmud and El-Sakka, Mahmoud R},
  journal={EURASIP journal on image and video processing},
  volume={2018},
  pages={1--12},
  year={2018},
  publisher={Springer}
}

@article{kermany2018identifying,
  title={Identifying medical diagnoses and treatable diseases by image-based deep learning},
  author={Kermany, Daniel S and Goldbaum, Michael and Cai, Wenjia and Valentim, Carolina CS and Liang, Huiying and Baxter, Sally L and McKeown, Alex and Yang, Ge and Wu, Xiaokang and Yan, Fangbing and others},
  journal={cell},
  volume={172},
  number={5},
  pages={1122--1131},
  year={2018},
  publisher={Elsevier}
}

\clearpage
% \appendix
\label{appendix:a}

\section*{Appendix A}

\begin{table}[h]
\centering
\scriptsize % Smaller font size
\caption{\textbf{Summary of IQA Metrics for CURE-OR Dataset Contrast Challenge  }}
\label{tab:IQAMetricsContrast}
\begin{adjustbox}{width=0.48\textwidth}
\begin{tabular}{@{}lS[table-format=2.6]S[table-format=1.3]S[table-format=1.3]S[table-format=1.3]S[table-format=1.3]S[table-format=1.3]S[table-format=1.3]S[table-format=1.0]@{}}
\toprule
\textbf{Noise level} & {\textbf{PSNR}} & {\textbf{SSIM}} & {\textbf{CW\_SSIM}} & {\textbf{UNIQUE}} & {\textbf{MS\_UNIQUE}} & {\textbf{CSV}} & {\textbf{SUMMER}} & {\textbf{F1 Score}} \\
\midrule
\multicolumn{8}{c}{\textbf{NBNET}} \\ % Centered bold label
\midrule

\textbf{Level 1}& 17.017 & 0.689 & 0.469 & 0.321 & 0.489 & 0.998 & 4.986& 0 \\
\textbf{Level 2}& 12.737 & 0.603 & 0.433 & 0.167 & 0.378 & 0.998 & 4.995& 0 \\
\textbf{Level 3}& 11.766 & 0.572 & 0.358 & 0.190 & 0.382 & 0.997 & 4.999& 0 \\
\textbf{Level 4}& 12.819 & 0.595 & 0.407 & 0.153 & 0.363 & 0.998 & 4.997& 0 \\
\textbf{Level 5}& 9.984& 0.500& 0.357& 0.000& 0.003& 0.997& 4.999& 0 \\
\midrule
\multicolumn{8}{c}{\textbf{BM3D}} \\ % Centered bold label
\midrule
\textbf{Level 1 } & 4.922 & 0.009 & 0.022 & 0.020 & 0.161 & 0.997 & 5.000    & 0.566 \\
\textbf{Level 2 } & 4.924 & 0.008 & 0.023 & 0.022 & 0.166 & 0.997 & 5.000    & 0 \\
\textbf{Level 3 } & 4.924 & 0.008 & 0.022 & 0.024 & 0.167 & 0.997 & 5.000    & 0 \\
\textbf{Level 4 } & 4.923 & 0.008 & 0.021 & 0.026 & 0.169 & 0.997 & 5.000    & 0 \\
\textbf{Level 5 } & 4.923 & 0.008 & 0.021 & 0.027 & 0.170 & 0.997 & 5.000    & 0 \\
\bottomrule
\end{tabular}
\end{adjustbox}
\end{table}

\begin{table}[h]
\centering
\scriptsize % Smaller font size
\caption{\textbf{Summary of IQA Metrics for Cure-TSR Dataset}}
\label{tab:IQAMetrics}
\begin{adjustbox}{width=0.48\textwidth}
\begin{tabular}{@{}lS[table-format=2.2]S[table-format=1.3]S[table-format=1.3]S[table-format=1.3]S[table-format=1.3]S[table-format=1.3]S[table-format=1.3]@{}}
\toprule
\textbf{Method} & {\textbf{PSNR}} & {\textbf{SSIM}} & {\textbf{CW\_SSIM}} & {\textbf{UNIQUE}} & {\textbf{MS\_UNIQUE}} & {\textbf{CSV}} & {\textbf{SUMMER}} \\
\midrule
\midrule
\multicolumn{8}{c}{\textbf{NBNET}} \\ % Centered bold label
\midrule

\textbf{Blur-1}& 19.231 & 0.870 & 0.295 & 0.273 & 0.534& 0.998 & 4.127 \\
\textbf{Blur-2}         &18.102& 0.850& 0.255&0.220&0.478&0.998&3.453\\
\textbf{Blur-3}         &15.684& 0.814& 0.331&0.121&0.376&0.998&4.791\\
\textbf{Blur-4}         &14.191& 0.788& 0.340&0.075&0.300&0.998&4.926\\
\textbf{Blur-5}         &13.297& 0.771& 0.344&0.049&0.253&0.998&4.958\\
\textbf{CodecError-1}        &11.948& 0.690& 0.224&0.004&0.123&0.997&4.069\\
\textbf{CodecError-2}        &11.712& 0.683& 0.225&0.003&0.113&0.997&4.038\\
\textbf{CodecError-3}        &11.070& 0.668& 0.232&0.002&0.096&0.997&4.305\\
\textbf{CodecError-4}        &10.593& 0.660& 0.237&0.001&0.083&0.997&4.439\\
\textbf{CodecError-5}        &10.194& 0.649& 0.241&0.001&0.075&0.997&4.481\\
\textbf{Darkening-1}    &30.895& 0.981& 0.342&0.855&0.903&0.999&4.954\\
\textbf{Darkening-2}    &10.467& 0.783& 0.383&0.760&0.849&0.998&4.998\\
\textbf{Darkening-3}    &6.9314& 0.474& 0.400&0.607&0.769&0.998&4.999\\
\textbf{Darkening-4}    &5.4872& 0.258& 0.397&0.468&0.673&0.998&4.999\\
\textbf{Darkening-5}    &4.8324& 0.135& 0.396&0.296&0.529&0.998&4.999\\
\textbf{Decolorizatio-1}  &28.405& 0.962& 0.297&0.741&0.827&0.998&4.538\\
\textbf{Decolorizatio-2}  &7.8521& 0.600& 0.379&8.928&0.027&0.997&4.794\\
\textbf{Decolorizatio-3}  &7.9170& 0.604& 0.380&9.595&0.027&0.997&4.818\\
\textbf{Decolorizatio-4}  &7.9899& 0.609& 0.378&0.000&0.027&0.997&4.840\\
\textbf{Decolorizatio-5}  &8.0628& 0.613& 0.366&0.000&0.028&0.997&4.847\\
\textbf{DirtyLens-1}    &32.352& 0.981& 0.348&0.854&0.903&0.999&4.988\\
\textbf{DirtyLens-2}    &27.814& 0.975& 0.347&0.830&0.888&0.999&4.978\\
\textbf{DirtyLens-3}    &20.825& 0.939& 0.362&0.736&0.833&0.998&4.981\\
\textbf{DirtyLens-4}    &17.075& 0.891& 0.375&0.578&0.743&0.998&4.980\\
\textbf{DirtyLens-5}    &13.888& 0.822& 0.394&0.326&0.553&0.998&4.993\\
\textbf{Exposure-1}     &24.944& 0.956& 0.334&0.634&0.782&0.998&4.955\\
\textbf{Exposure-2}     &15.841& 0.863& 0.328&0.278&0.557&0.998&4.868\\
\textbf{Exposure-3}     &11.506& 0.753& 0.330&0.082&0.349&0.998&4.802\\
\textbf{Exposure-4}     &9.1403& 0.687& 0.333&0.015&0.201&0.997&4.828\\
\textbf{Exposure-5}     &8.0153& 0.663& 0.334&0.001&0.118&0.997&4.893\\
\textbf{Haze-1}         &27.822& 0.966& 0.313&0.753&0.840&0.998&4.581\\
\textbf{Haze-2}         &19.835& 0.916& 0.365&0.788&0.867&0.998&4.993\\
\textbf{Haze-3}         &14.513& 0.833& 0.376&0.651&0.790&0.998&4.995\\
\textbf{Haze-4}         &6.7585& 0.469& 0.280&0.206&0.324&0.997&3.638\\
\textbf{Haze-5}         &6.8823& 0.520& 0.287&0.219&0.395&0.997&4.349\\
\textbf{LensBlur-1}     &21.269& 0.904& 0.297&0.389&0.620&0.998&3.889\\
\textbf{LensBlur-2}     &17.296& 0.838& 0.334&0.177&0.442&0.998&4.625\\
\textbf{LensBlur-3}     &14.812& 0.789& 0.345&0.090&0.313&0.998&4.894\\
\textbf{LensBlur-4}     &13.592& 0.767& 0.350&0.052&0.254&0.998&4.958\\
\textbf{LensBlur-5}     &12.837& 0.752& 0.351&0.030&0.214&0.998&4.978\\
\textbf{Noise-1}        &23.903& 0.937& 0.291&0.561&0.724&0.998&4.115\\
\textbf{Noise-2}        &24.497& 0.934& 0.316&0.543&0.732&0.998&4.674\\
\textbf{Noise-3}        &20.767& 0.891& 0.320&0.358&0.615&0.998&4.771\\
\textbf{Noise-4}        &17.700& 0.842& 0.318&0.213&0.493&0.998&4.722\\
\textbf{Noise-5}        &15.716& 0.802& 0.317&0.129&0.397&0.998&4.686\\
\textbf{Rain-1}         &21.547& 0.919& 0.298&0.587&0.746&0.998&4.817\\
\textbf{Rain-2}         &19.413& 0.897& 0.310&0.518&0.706&0.998&4.871\\
\textbf{Rain-3}         &18.722& 0.888& 0.312&0.486&0.686&0.998&4.899\\
\textbf{Rain-4}         &18.291& 0.883& 0.314&0.467&0.673&0.998&4.909\\
\textbf{Rain-5}         &17.305& 0.872& 0.324&0.340&0.589&0.998&4.963\\
\textbf{Shadow-1}       &25.945& 0.959& 0.290&0.710&0.809&0.998&4.528\\
\textbf{Shadow-2}       &20.935& 0.943& 0.304&0.678&0.788&0.998&4.759\\
\textbf{Shadow-3}       &18.126& 0.896& 0.311&0.588&0.730&0.998&4.757\\
\textbf{Shadow-4}       &16.127& 0.816& 0.320&0.496&0.663&0.998&4.750\\
\textbf{Shadow-5}       &14.647& 0.708& 0.322&0.405&0.586&0.998&4.709\\
\textbf{Snow-1}         &29.615& 0.969& 0.308&0.782&0.859&0.999&4.922\\
\textbf{Snow-2}         &21.488& 0.921& 0.276&0.522&0.704&0.998&4.691\\
\textbf{Snow-3}         &17.319& 0.870& 0.268&0.341&0.582&0.998&4.473\\
\textbf{Snow-4}         &13.780& 0.804& 0.276&0.180&0.455&0.998&4.391\\

\bottomrule
\end{tabular}
\end{adjustbox}
\end{table}

\begin{table}[h]
\centering
\scriptsize % Smaller font size
\captionsetup{font=small}
\caption{\textbf{Summary of IQA Metrics for Grayscale Salt and Pepper Challenge}}
\label{tab:GrayscaleSaltPepper}
\begin{adjustbox}{width=0.48\textwidth}
\begin{tabular}{@{}lS[table-format=2.6]S[table-format=1.3]S[table-format=1.3]S[table-format=1.3]S[table-format=1.3]S[table-format=1.3]S[table-format=1.3]S[table-format=1.0]@{}}
\toprule
\textbf{Noise level} & {\textbf{PSNR}} & {\textbf{SSIM}} & {\textbf{CW\_SSIM}} & {\textbf{UNIQUE}} & {\textbf{MS\_UNIQUE}} & {\textbf{CSV}} & {\textbf{SUMMER}} & {\textbf{F1 Score}} \\
\midrule
\multicolumn{8}{c}{\textbf{NBNET}} \\ % Centered bold label
\midrule
\textbf{Level 1} & 19.508 & 0.485 & 0.524 & 0.135 & 0.298 & 0.998 & 4.927 & 0 \\
\textbf{Level 2} & 16.624 & 0.367 & 0.458 & 0.075 & 0.205 & 0.998 & 4.886 & 0 \\
\textbf{Level 3} & 17.755 & 0.592 & 0.542 & 0.152 & 0.361 & 0.998 & 4.882 & 0 \\
\textbf{Level 4} & 13.682 & 0.470 & 0.525 & 0.077 & 0.269 & 0.997 & 4.739 & 0 \\
\textbf{Level 5} & 10.885 & 0.247 & 0.546 & 0.000 & 0.051 & 0.997 & 4.661 & 0 \\
\midrule
\multicolumn{8}{c}{\textbf{BM3D}} \\ % Centered bold label
\midrule
\textbf{Level 1}& 26.485 & 0.841 & 0.792 & 0.555& 0.732 & 0.998 & 4.985& 0.406 \\
\textbf{Level 2}& 21.339 & 0.584 & 0.701 & 0.351& 0.579 & 0.998 & 4.977& 0 \\
\textbf{Level 3}& 17.760 & 0.714 & 0.555 & 0.232& 0.479 & 0.998 & 4.938& 0 \\
\textbf{Level 4} & 15.374 & 0.642 & 0.423 & 0.136& 0.368 & 0.998 & 4.925& 0 \\
\textbf{Level 5} & 13.292 & 0.486 & 0.293 & 0.021& 0.188 & 0.997 & 4.916& 0 \\
\bottomrule
\end{tabular}
\end{adjustbox}
\end{table}

\begin{table}[h]
\centering
\scriptsize % Smaller font size
\caption{\textbf{Summary of IQA Metrics for Cure-TSR Dataset}}
\label{tab:IQAMetrics}
\begin{adjustbox}{width=0.48\textwidth}
\begin{tabular}{@{}lS[table-format=2.2]S[table-format=1.3]S[table-format=1.3]S[table-format=1.3]S[table-format=1.3]S[table-format=1.3]S[table-format=1.3]@{}}
\toprule
\textbf{Method} & {\textbf{PSNR}} & {\textbf{SSIM}} & {\textbf{CW\_SSIM}} & {\textbf{UNIQUE}} & {\textbf{MS\_UNIQUE}} & {\textbf{CSV}} & {\textbf{SUMMER}} \\
\midrule
\midrule
\multicolumn{8}{c}{\textbf{BM3D}} \\ % Centered bold label
\midrule

\textbf{Blur-1			}&			20.918& 	0.761	&0.442	&   	0.406&	    0.527&	   	0.998&	4.997\\	
\textbf{Blur-2			}&			19.088& 	0.679	&0.345	&    	0.270&	   	0.408&	    0.998&	4.997\\	
\textbf{Blur-3			}&			16.212& 	0.491	&0.202	&   	0.071&	   	0.208&	   	0.998&	4.997\\	
\textbf{Blur-4			}&			15.865& 	0.446	&0.198	&   	0.074&	   	0.187&	    0.998&	4.998\\	
\textbf{Blur-5			}&			14.691& 	0.369	&0.153	&   	0.020&	   	0.118&	   	0.997&	4.998\\	
\textbf{CodecError-1	}&			13.157& 	0.337	&0.519	&   	0.045&	   	0.108&	   	0.999&	4.950\\	
\textbf{CodecError-2	}&			13.042& 	0.319	&0.517	&   	0.047&	   	0.110&	   	0.997&	4.945\\	
\textbf{CodecError-3	}&			12.286& 	0.241	&0.486	&    	0.027&	   	0.070&	   	0.999&	4.959\\	
\textbf{CodecError-4	}&			11.936& 	0.210	&0.464	&   	0.023&	   	0.064&	   	0.997&	4.967\\	
\textbf{CodecError-5	}&			11.701& 	0.180	&0.439	&    	0.020&	   	0.058&	   	0.997&	4.972\\	
\textbf{Darkening-1		}&			16.322& 	0.676	&0.411	&   	0.419&	   	0.544&	   	0.998&	4.998\\	
\textbf{Darkening-2		}&			10.431& 	0.288	&0.146	&   	0.160&	   	0.288&	   	0.998&	4.999\\	
\textbf{Darkening-3		}&			8.4042& 	0.105&0.053	&   	0.050&	   	0.154&	   	0.997&	4.999\\	
\textbf{Darkening-4		}&			7.4923& 	0.038	&0.026	&   	0.020&	   	0.103&	   	0.997&	4.999\\	
\textbf{Darkening-5		}&			7.0206& 	0.012	&0.013	&   	0.006&	   	0.071&	   	0.997&	4.999\\	
\textbf{Decolorization-1}&			23.208& 	0.810	&0.585	&   	0.534&	   	0.645&	   	0.999&	4.997\\	
\textbf{Decolorization-2}&			22.004& 	0.761&0.582	&   	0.500&	   	0.621&	   	0.999&	4.998\\	
\textbf{Decolorization-3}&			20.685& 	0.679	&0.578	&   	0.444&	   	0.585&	   	0.998&	4.998\\	
\textbf{Decolorization-4}&			19.460& 	0.569	&0.575	&   	0.358&	   	0.538&	   	0.998&	4.999\\	
\textbf{Decolorization-5}&			18.583& 	0.471	&0.570	&   	0.282&	   	0.495&	   	0.998&	4.999\\	
\textbf{DirtyLens-1		}&			23.221& 	0.813	&0.581	&   	0.531&	   	0.642&	   	0.999&	4.997\\	
\textbf{DirtyLens-2		}&			22.212& 	0.818	&0.605	&   	0.542&	   	0.651&	   	0.999&	4.996\\	
\textbf{DirtyLens-3		}&			15.032& 	0.504	&0.222	&   	0.118&	   	0.245&	   	0.998&	4.994\\	
\textbf{DirtyLens-4		}&			13.286& 	0.437	&0.183	&       0.088&	   	0.210&	   	0.998&	4.995\\	
\textbf{DirtyLens-5		}&			12.357& 	0.289	&0.136	&   	0.049&	   	0.161&	   	0.997&	4.998\\	
\textbf{Exposure-1		}&			18.930& 	0.848	&0.849	&   	0.617&	   	0.699&	   	0.999&	4.993\\	
\textbf{Exposure-2		}&			12.269& 	0.578	&0.730	&   	0.290&	   	0.412&	    0.998&	4.969\\	
\textbf{Exposure-3		}&			9.2365& 	0.371	&0.617	&   	0.080&	   	0.200&	   	0.997&	4.952\\	
\textbf{Exposure-4		}&			7.5998& 	0.219	&0.488	&   	0.014&	    0.090&	   	0.997&	4.947\\	
\textbf{Exposure-5		}&			6.9609& 	0.142	&0.393	&   	0.003&	   	0.060&	   	0.997&	4.956\\	
\textbf{Haze-1			}&			17.632& 	0.863	&0.876	&   	0.858&	    0.883&	   	0.999&	4.998\\	
\textbf{Haze-2			}&			13.602& 	0.716&0.770	&   	0.775&	   	0.801&	   	0.998&	4.999\\	
\textbf{Haze-3			}&			10.701& 	0.502	&0.586	&   	0.584&	   	0.633&	   	0.998&	4.999\\	
\textbf{Haze-4			}&			6.3609& 	0.311	&0.361	&   	0.396&	   	0.455&	   	0.998&	4.999\\	
\textbf{Haze-5			}&			8.2096& 	0.242	&0.269	&   	0.232&	   	0.311&	   	0.998&	4.999\\	
\textbf{LensBlur-1		}&			23.110& 	0.858	&0.602	&   	0.576&	   	0.667&	   	0.999&	4.997\\	
\textbf{LensBlur-2		}&			18.918& 	0.665&0.389	&   	0.226&	   	0.366&	   	0.998&	4.997\\	
\textbf{LensBlur-3		}&			16.984& 	0.532	&0.313	&   	0.092&	    0.222&	    0.998&	4.997\\	
\textbf{LensBlur-4		}&			15.918& 	0.455	&0.257	&   	0.044&	   	0.154&	   	0.998&	4.997\\	
\textbf{LensBlur-5		}&			15.226& 	0.405	&0.220	&   	0.024&	   	0.116&	   	0.997&	4.997\\	
\textbf{Noise-1			}&			23.426& 	0.820	&0.583	&   	0.520&	   	0.636&	    0.999&	4.997\\	
\textbf{Noise-2			}&			22.167& 	0.778	&0.557	&   	0.440&	   	0.575&	   	0.999&	4.996\\	
\textbf{Noise-3			}&			19.906& 	0.693	&0.490	&   	0.315&	   	0.468&	   	0.998&	4.995\\	
\textbf{Noise-4			}&			17.686&	    0.580	&0.399	&   	0.201&	   	0.358&	   	0.998&	4.993\\	
\textbf{Noise-5			}&			15.416& 	0.418	&0.201	&   	0.083&	   	0.211&	   	0.998&	4.994\\	
\textbf{Rain-1			}&			14.845& 	0.404	&0.214	&   	0.164&	   	0.294&	   	0.998&	4.999\\	
\textbf{Rain-2			}&			13.550& 	0.315	&0.156	&   	0.107&	   	0.226&	    0.997&	4.999\\	
\textbf{Rain-3			}&			13.211& 	0.292	&0.140	&   	0.091&	   	0.207&	   	0.997&	4.999\\	
\textbf{Rain-4			}&			12.900& 	0.272&0.129	&   	0.081&	   	0.194&	   	0.997&	4.999\\	
\textbf{Rain-5			}&			12.646& 	0.222	&0.105	&   	0.049&	   	0.154&	   	0.997&	4.999\\	
\textbf{Shadow-1		}&			21.875& 	0.943	&0.902	&   	0.818&	   	0.857&	   	1.001&	4.993\\	
\textbf{Shadow-2		}&			20.599& 	0.928	&0.889	&   	0.698&	   	0.771&	   	0.999&	4.993\\	
\textbf{Shadow-3		}&			15.340& 	0.832	&0.847	&   	0.580&	    0.666&	   	0.998&	4.953\\	
\textbf{Shadow-4		}&			10.379& 	0.662	&0.781	&       0.480&	   	0.565&	   	0.998&	4.651\\	
\textbf{Shadow-5		}&			7.4166& 	0.566&0.731	&   	0.421&	   	0.502&	   	0.997&	4.612\\	
\textbf{Snow-1			}&			22.141& 	0.924	&0.899	&   	0.823&	   	0.866&	   	0.999&	4.998\\	
\textbf{Snow-2			}&			20.664& 	0.840	&0.813	&   	0.530&	   	0.644&	   	0.999&	4.991\\	
\textbf{Snow-3			}&			18.965& 	0.724	&0.771	&   	0.321&	   	0.477&	   	0.998&	4.995\\	
\textbf{Snow-4			}&			15.726& 	0.579	&0.726	&   	0.180&	   	0.332&	   	0.998&	4.993\\	

\bottomrule
\end{tabular}
\end{adjustbox}
\end{table}

 % use "AppendixA.tex" as the filename (no space)

% \clearpage 

\end{document}